\documentclass[lettersize,journal]{IEEEtran}

\usepackage{amsmath,amsfonts}
\usepackage{algorithmic}
\usepackage{algorithm}
\usepackage{array}
\usepackage[caption=false,font=normalsize,labelfont=sf,textfont=sf]{subfig}
\usepackage{textcomp}
\usepackage{stfloats}
\usepackage{url}
\usepackage{verbatim}
\usepackage{graphicx}
\usepackage{cite}
\usepackage[bottom]{footmisc}
\usepackage{mathptmx}
\usepackage{graphicx}
\usepackage{siunitx}
\usepackage{times}
\usepackage[T1]{fontenc}
\usepackage{enumitem}
\usepackage[table]{xcolor} % loads also »colortbl«
\usepackage{wrapfig}
\usepackage{multirow}
\usepackage{adjustbox}
\usepackage{mwe}
\usepackage{booktabs}
\usepackage{graphbox}
\usepackage{dirtytalk}
\usepackage{soul}
\usepackage{mdframed}
\usepackage{tabularx}
\usepackage{xcolor}
\usepackage{enumitem}
\usepackage{xspace}
\usepackage{tcolorbox}
\usepackage{color}
\usepackage{booktabs} % For professional looking tables
\usepackage{caption}  % For custom captions
\usepackage{arydshln}

\definecolor{CindySalmon}{RGB}{232, 125, 114}
\definecolor{greenB}{RGB}{77, 175, 74}
\definecolor{purpleF}{RGB}{152,78,163}

\definecolor{RoyalBlue}{HTML}{4169E1}
\newcommand{\re}[1]{#1}

\usepackage[bookmarks,backref=true,linkcolor=black]{hyperref}

\begin{document}

\title{Do You ``Trust'' This Visualization? \\ An Inventory to Measure Trust in Visualizations}

\author{%
Huichen Will Wang, Kylie Lin, Andrew Cohen, Ryan Kennedy, Zach Zwald, Carolina Nobre, \\Cindy Xiong Bearfield
\thanks{Huichen Will Wang did this work as a research assistant for Cindy Xiong Bearfield at the University of Massachusetts Amherst. He is now with the University of Washington. E-mail: wwill@cs.washington.edu.}
\thanks{Kylie Lin and Cindy Xiong Bearfield are with Georgia Tech. E-mail: \{klin368, cxiong\}@gatech.edu.}
\thanks{Andrew Cohen is with the University of Massachusetts Amherst. E-mail: alc@umass.edu.}
\thanks{Ryan Kennedy is with Ohio State University. E-mail: kennedy.310@osu.edu.}
\thanks{Zach Zwald is with the University of Houston. E-mail: zjzwald@central.uh.edu.}
\thanks{Carolina Nobre is with the University of Toronto. E-mail: cnobre@cs.toronto.edu.}}
% \thanks{Cindy Xiong Bearfield is with Georgia Tech. E-mail: cxiong@gatech.edu.}}

\markboth{Journal of \LaTeX\ Class Files,~Vol.~14, No.~8, August~2021}%
{Shell \MakeLowercase{\textit{et al.}}: A Sample Article Using IEEEtran.cls for IEEE Journals}


\maketitle

\begin{figure*}[!t]
  \centering
  \includegraphics[width = \textwidth]{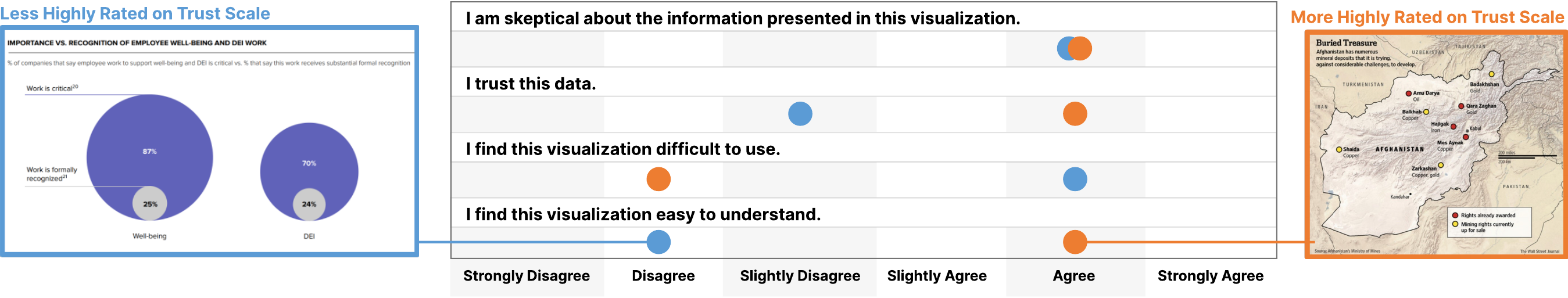}
  \caption{The four visualization-related questions in our inventory measure trust in visualizations by capturing visualization clarity, usability, data credibility, and information comprehensibility. The map on the right is rated as more trustworthy because participants perceive its data as more reliable, find it easier to use, and consider it more comprehensible.}
  \label{fig:baseline}
\end{figure*}

\begin{abstract}
Trust plays a critical role in visual data communication and decision-making, yet existing visualization research employs varied trust measures, making it challenging to compare and synthesize findings across studies. In this work, we first took a bottom-up, data-driven approach to understand what visualization readers mean when they say they ``trust'' a visualization.
We compiled and adapted a broad set of trust-related statements from existing inventories and collected responses to visualizations with varying degrees of trustworthiness. Through exploratory factor analysis, we derived an operational definition of trust in visualizations. Our findings indicate that people perceive a trustworthy visualization as one that presents credible information and is comprehensible and usable. Building on this insight, we developed an \re{eight-item} inventory\re{: four core items measuring trust in visualizations and four optional items controlling for individual differences in baseline trust tendency.} We \re{established the inventory's internal consistency reliability using McDonald's omega, confirmed its content validity by demonstrating alignment with theoretically-grounded trust dimensions, and validated its criterion validity} through two trust games with real-world stakes. Finally, we illustrate how this standardized inventory can be applied across diverse visualization research contexts. Utilizing our inventory, future research can examine how design choices, tasks, and domains influence trust, and how to foster appropriate trusting behavior in human-data interactions.
\end{abstract}

\begin{IEEEkeywords}
survey measurements, behavioral measurements, investment game, trust, visualizations, validation.
\end{IEEEkeywords}

\section{Introduction}

\IEEEPARstart{P}{eople} increasingly rely on visualizations to make decisions~\cite{9552846,oral2023information}. While calibrated trust supports informed decision-making, misplaced trust can lead to misinformation and missteps. Despite the importance of trust in visualizations, it remains a loosely characterized concept~\cite{elhamdadi2022trust}. In this work, we adopted a bottom-up approach and examined what readers mean when they say they ``trust'' a visualization. Through this data-driven investigation, we identified the underlying factors shaping viewer trust and established an operational definition for it. Building on this definition, we developed and validated a \re{compact eight-item inventory: four core items measuring trust in visualizations and four optional items measuring general trust disposition to control for individual differences in trusting tendency (Table~\ref{tab:trust_inventory})}.

Traditionally, the visualization community has relied on single-item Likert scales to measure trust, often in the form of ``on a scale from 1 to \textit{N}, how much do you trust this visualization?''~\cite{padilla2022multiple, kim2020bayesian, zhou2019effects}. However, trust is a \textit{nuanced and multifaceted} construct. Existing work in social sciences has identified at least three factors associated with trustworthiness: ability, benevolence, and integrity~\cite{mayer1995integrative, zheng2002trust, uslaner2018oxford}. 
Psychology research has further shown that a person's tendency to trust is a stable personality trait~\cite{evans2008survey}, which implies that individual differences should also be controlled for when collecting trust ratings.
These findings call into question the validity of measuring trust in visualizations with a simplistic, one-dimensional statement that likely inadequately represents the full complexity of the construct and invites different interpretations.

Recognizing this complexity, visualization researchers have adopted varied definitions and operationalizations that emphasize different dimensions of trust. For example, Kong et al.~\cite{kong2019trust} define visualization trustworthiness as perceived credibility and appropriateness, while Burns et al. ~\cite{burns2022invisible} focus on persuasiveness and transparency. Taking a behavioral approach, Zehrung et al.~\cite{zehrung2021vis} and Xiong et al.~\cite{xiong2019examining} treat people's preference for using a visualization (system) as a proxy for trust. These diverse operationalizations of trust in visualization research have, in turn, made it difficult to compare and synthesize study outcomes. 

% Futher, are these measures of trust across studies even comparable, and if so, to what extent?
% We raise the question of wehtehr reader trust in a visualization is something that can be captured in a single survey item, given the multi-facetedness of trust, and posit there should not  be a \textit{single} definition for trust in visualizations. 
% So how might we reconcile the need to capture the multi-faceted-ness of trust while also elicit trust in experiments in a way that enables researchers to compare effects and synthesize outcomes across multiple studies?

\begin{tcolorbox}
How can we establish a definition that captures the multifaceted nature of trust and develop a standardized inventory for its measurement?
\end{tcolorbox}

We address this challenge through a bottom-up, data-driven approach. We sought to understand what visualization readers mean when they say they ``trust'' a visualization, allowing the definition of trust to emerge from empirical data rather than imposing predetermined conceptualizations. Doing so enables us to understand the dimensions of trust as they are actually experienced and interpreted by readers. 
% We navigate this trade-off through a data-driven approach, examining what readers mean when they ``trust'' a visualization and curating a trust inventory based on the most diagnostic factors influencing their reported trust.

We began by compiling trust-eliciting statements from existing inventories across multiple disciplines and adapting them to the visualization context.
To account for the possibility that different phrasings in the adaptation might be perceived differently by the reader, we created multiple versions of the same statements as appropriate.
%, as shown in Tables~\ref{trustQuestions1} and ~\ref{trustQuestions2}.
By collecting participant responses on visualizations with different levels of trustworthiness, we revealed with Exploratory Factor Analysis (EFA) which items clustered together (measuring the same facet of trust) and which items failed to load onto any factors (potentially less relevant to trust). Through interpreting the resulting factors, we found that people consider a trustworthy visualization as one that presents credible information and is comprehensible and usable.

% We then conducted an Exploratory Factor Analysis (EFA) on their responses to uncover factors they considered the most when reporting on their trust in a visualization. 

% the underlying dimensions of trust in visualizations and determine the statements that most diagnostically measure each dimension. 

We then built on this operational definition to develop the trust inventory. By examining factor loadings and item discriminability scores, we condensed the initial pool of items into \re{a compact inventory consisting of two components}---four visualization-related \re{questions} and four \re{questions} that allow researchers to control for individual trust dispositions. Figure~\ref{fig:baseline} shows the four visualization-related questions in our trust inventory \re{applied to two visualizations. Table~\ref{tab:trust_inventory} shows all eight items and the prompts we recommend researchers to adopt.} These items distinctly cover the important dimensions of trust that participants considered when evaluating data visualizations, including measurements of visualization clarity, usability, data credibility, and information comprehensibility. Overall, this compact inventory provides a standardized yet nuanced approach to measuring trust in visualizations. 

\re{To ensure our inventory is both reliable and valid, we conducted multiple assessments. We established the inventory's \textbf{internal consistency reliability} through McDonald's omega, showing that items within each component consistently measure their respective constructs. We evaluated \textbf{content validity} by ensuring our items accurately and comprehensively cover established trust dimensions---usability, comprehensibility, and credibility---identified in prior visualization trust research. Finally, we assessed \textbf{criterion validity} through validation studies using an adapted trust game across two different contexts. The results confirmed that our trust measures effectively predict behavioral outcomes, demonstrating the inventory's practical utility for visualization research.}

In sum, our primary contributions are as follows:
\begin{itemize}
    \item A factor-analysis-powered investigation revealing that \re{viewers perceive a trustworthy visualization as one that presents credible information and is comprehensible and usable;}
    %readers' trust in visualizations is primarily shaped by data credibility, visualization clarity, and the reader's general trust dispositions;
    \item A compact inventory consisting of statements that effectively represent each trust factor and exhibit high item discrimination;
    \item \re{Careful validation of the inventory, establishing its internal consistency reliability (sufficient McDonald's omega), content validity (alignment with established trust frameworks), and criterion validity (through trust games confirming our measures effectively predict behavioral trust);}
    %Validation of the inventory through investment games, demonstrating that our survey measures effectively predict behavioral trust;
    \item Three representative application scenarios illustrating how our inventory can be applied in visualization research to study trust in different contexts.
\end{itemize}

\begin{table}[h!]
\centering
\caption{\re{Our full, validated inventory consists of four items measuring trust in visualizations and four items measuring general trust disposition. The latter four are optional and help control for baseline trusting tendency. All items are measured on a six-point Likert scale (Strongly Disagree, Disagree, Slightly Disagree, Slightly Agree, Agree, and Strongly Agree).}}
\label{tab:trust_inventory}
\begin{tabular}{p{0.9\linewidth}}
\toprule
\textbf{\re{Trust in Visualization}} \\
\textit{\re{Indicate the extent to which you agree with each statement:}}\\
\noalign{\vskip\aboverulesep}
\hdashline
\noalign{\vskip\belowrulesep}
\quad \re{I find this visualization easy to understand.} \\
\quad \re{I find this visualization difficult to use.} \\
\quad \re{I am skeptical about the information presented in this visualization.} \\
\quad \re{I trust this data.} \\
\midrule
\textbf{\re{General Trust Disposition}} \\
\textit{\re{Indicate the extent to which each statement describes you:}} \\
\noalign{\vskip\aboverulesep}
\hdashline
\noalign{\vskip\belowrulesep}
\quad \re{Have a good word for everyone} \\
\quad \re{Retreat from others} \\
\quad \re{Avoid contacts with others} \\
\quad \re{Respect others} \\
\bottomrule
\end{tabular}
\end{table}

\section{Related Work}
\label{section:rw}

While extensive research has examined how different aspects of a visualization influence viewer perception, a systematic exploration of their relationship to trust remains lacking \cite{elhamdadi2022trust}. In this section, we review existing theories on trust from the social sciences and the visualization community, along with the methods used to measure trust in these contexts\re{, which together inform the development of our inventory for measuring trust in visualizations.}

%%%%%%%%%%%%%%%%%%%%%%%%%%%%%%%%%%%%%%%%%%%%%%
\subsection{Trust in the Social Sciences}
Trust in the social sciences is commonly defined through four factors proposed by McKnight and Chervany: benevolence (morality and responsiveness), competence (ability to achieve goals), integrity (honesty and reliability), and predictability (consistent behavior)~\cite{mcknight2000trust}. However, adapting this definition to measure trust in technology has proven somewhat difficult. Some scholars have argued that people consider machines as ``mere tools,'' where reliability and predictability of behavior are sufficient for them to ``trust'' the system~\cite{roff2018trust}. In essence, this viewpoint advocates for keeping the competence and predictability components of McKnight and Chervany's typology, but discarding the ``interpersonal'' elements of benevolence and integrity. This definition of trust also follows from the instrumental logic of game theoretic rational choice decision-making, wherein a human trusts the machine to the extent that they conclude doing so will further their self-interest~\cite{kirkpatrick2016can}.

Others have argued, however, that benevolence and integrity cannot be disregarded, even in the context of technology systems. Lee and See~\cite{lee2004trust}, for example, conceptualize integrity as ``the degree to which the trustee adheres to a set of principles the trustor finds acceptable,'' and benevolence as ``the extent to which the intents and motivations of the trustee are aligned with those of the trustor.'' From their perspective, people can infer these characteristics in technology. Similarly, in the context of artificial intelligence (AI), Roff and Danks~\cite{roff2018trust} advance that interpersonal trust requires that the human views the AI as a ``moral agent with values and preferences.'' In this work, we investigate trust in an important product of technology systems---data visualizations. Instead of imposing a definition on trust in visualizations, we seek to understand this construct from the reader's perspective and identify statements that diagnostically elicit different aspects of trust.

%%%%%%%%%%%%%%%%%%%%%%%%%%%%%%%%%%%%%%%%%%%%%%
\subsection{Theorizing Trust in Data Visualizations}

Perceived trustworthiness of visualizations and visualization systems can change how people conduct data analysis and make decisions \cite{honeycutt2020soliciting, nourani2019effects, hullman2011visualization, dork2013critical, gaba2023my}. 
For example, when scientists explore unfamiliar domains, they report model outputs communicated via a specialized tool to be more trustworthy than general tools such as Excel or MATLAB \cite{dasgupta2016familiarity}.
Researchers have also identified the volume and detail of information in visualizations as predictors of trust and perceived transparency \cite{hood2006transparency, xiong2019examining}. 
While disclosing information about data uncertainty can improve trust \cite{jung2015displayed, joslyn2013, padilla2022multiple}, excessive information can overwhelm a viewer and distract from key takeaways \cite{ajani2021declutter, xiong2019curse}, ultimately reducing interpretability \cite{poursabzi2021manipulating} and trust \cite{elhamdadi2023vistrust}. 

To shed light on trust antecedents, researchers have recently developed frameworks for trust in data visualizations. Pandey et al.~\cite{pandey2023you} reviewed the literature and identified five factors related to trust in two studies: familiarity (how familiar the viewers are with the data and topic of the visualization), clarity (how understandable the visualization is), credibility (how authentic and accurate the data is), reliability (how well the visualization might support sound decisions), and confidence (how confident the viewers feel using the visualization to make decisions).
Elhamdadi et al.~\cite{elhamdadi2023vistrust} synthesized foundational theories and proposed that trust can be either cognitive or affective, and that it can be directed toward the data or the visualization design. Display clarity, display accuracy, display usability, data currency, data coverage, data accuracy, and data clarity contribute to cognitive trust, while display aesthetics, display appropriateness (i.e., whether a visualization is designed with deceptive intent), source credibility, and data soundness contribute to affective trust. In this work, we extend this line of research by taking a data-driven approach to explore readers' conception of trust in visualizations. In addition, we contribute a compact inventory consisting of statements that diagnostically measure dimensions of trust identified via factor analysis.

\subsection{Measuring Trust in Data Visualizations}
Computer scientists often use substitution variables as proxies for trust. For example, high usability and positive user experiences are correlated with trust \cite{mayr2019trust, costante2011line, sasse2005usability}, motivating Zehrung et al.~\cite{zehrung2021vis} and Xiong et al.~\cite{xiong2019examining} to operationalize trusting a visualization (system) as people's preference for using it. Further, perceived bias in visualizations has been used as a proxy measure of trust, with higher perceived bias linked to less trust~\cite{kong2019trust}. Additionally, the colorfulness and visual complexity of a website can impact its perceived usability~\cite{harrison2015infographic}, making them reasonable proxies for trust in websites~\cite{reinecke2013predicting}. However, in the context of visualizations, even beautifully designed visualizations with highlights and annotations may still lead readers to question the authors' motives, potentially decreasing trust~\cite{ajani2021declutter}. 

Another widely adopted tool for measuring trust in data visualizations is Likert scales. Some representative examples include ``on a scale from 1 to 7, how much do you trust this visualization?'' \cite{xiong2019examining}, ``on a scale of 1 to 100, how much do you trust that the base data is accurate?'' \cite{kim2017data}, and ``on a scale from 1 to 9, how much do you trust the visualized model predictions?'' \cite{zhou2019effects}. While easy to administer, these Likert scale questions have not been tested for reliability and validity. Further, assessing trust via a single question may invite varied interpretations that diverge from what the researcher intends to measure. In this work, we dissect what readers mean when they say they ``trust'' a visualization, contributing an important step towards establishing robust definitions and measurement tools for trust in data visualizations.

\section{Comprehensive Compilation of Trust-Related Items for Visualizations}
\label{inventoryItems}

To effectively identify a set of questions that capture different facets of trust in visualizations, we compiled a collection of trust-related items from prior research on trust theories and measurement tools across various contexts, such as online firms~\cite{bhattacherjee2002individual}, visualizations~\cite{xiong2019examining, pandey2023you}, organizations~\cite{mayer1995integrative}, and interpersonal relationships~\cite{evans2008survey}. For example, Mayer et al.~\cite{mayer1995integrative} identified three factors associated with trustworthiness: ability, benevolence, and integrity. Bhattacherjee et al.~\cite{bhattacherjee2002individual} considered the expression of these factors in the context of a person interacting with an online firm. Elhamdadi et al.~\cite{elhamdadi2023vistrust} categorized trust in visualizations by whether it was directed towards the underlying data or the visual design aesthetics and layout.
Pandey et al.~\cite{pandey2023you} additionally considered the visualization source and the visualization user's individual characteristics. 

We adapted trust-related items from these sources to align with the context of data visualizations. For example, we modified items in Bhattacherjee et al.~\cite{bhattacherjee2002individual} to focus on user perception of visualization creators rather than corporate entities. We also introduced new items to ensure broad coverage by crossing the key dimensions of trust identified in previous studies. For example, Elhamdadi et al.'s visualization trust framework~\cite{elhamdadi2023vistrust} suggests that trust in a visualization can be separately directed towards the visualization \textbf{creator}, the visualized \textbf{dataset}, and the visualization \textbf{design} choices. We created new items by crossing these dimensions with the three factors associated with trust identified by Mayer et al.~\cite{mayer1995integrative}: ability, benevolence, and integrity. \re{We provide a list of all pooled questions along with the dimensions they represent in Supplemental Materials.}
%The details of these trust questions can be found in Tables~\ref{trustQuestions1} and~\ref{trustQuestions2}.

To give another example, applying this framework to an item from Bhattacherjee et al.~\cite{bhattacherjee2002individual}, which originally stated, ``Amazon is fair in its conduct of customer transactions,'' we created variations focusing on fair collection, processing, and visualization of data. Specifically, for the \textit{creator} dimension, we developed three statements: (1) ``The creator is fair in its collection of data'' (\#67)\footnote{\re{Item numbers in parentheses (e.g., \#67) refer to assigned IDs in our initial pool of 113 trust-related items. All items with their corresponding IDs are provided in a table in the Supplemental Materials.}}, (2) ``The creator is fair in the process of creating this visualization'' (\#68), and (3) ``The creator of this visualization is unbiased and trustworthy'' (\#91). For the \textit{data} dimension, the modified statement read: ``The data in this visualization is unbiased and trustworthy'' (\#46). For the \textit{design} dimension, we used: ``I think this visualization depicts data in an objective manner'' (\#7).

We also included items that capture visualization-related behavioral outcomes and general trust dispositions, following Elhamdadi et al.'s trust framework~\cite{elhamdadi2023vistrust}. Our inventory encompasses five dimensions of trust: trust in design, trust in data, trust in creators, behavioral outcomes, and individual characteristics. To ensure response reliability, we included six pairs of reverse-worded items across all dimensions.
%, as indicated by the item pairs labeled with (R\#) in Tables~\ref{trustQuestions1} and ~\ref{trustQuestions2}. 
For example, within the \textit{Trust in Design} dimension, we included both a positively worded item, ``I like this visualization'' (\#17), and a negatively worded counterpart, ``I dislike this visualization'' (\#18). These reverse-worded items help identify inconsistent responses, enhancing data quality in downstream crowd-sourcing experiments.

Previous research has shown that even small variations in question phrasing can substantially affect viewer responses~\cite{rasinski1989effect, fan2010factors}. To account for the potential influence of phrasing on responses, we brainstormed variations of items that conveyed semantically similar meanings but were worded differently. For instance, in assessing the credibility of the \textit{creator(s)}, we included items such as ``The creator(s) of the visualization is respected'' (\#78) and ``The creator(s) of the visualization has a good reputation'' (\#79). These variations aimed to capture finer-grained angles of trust. Rather than selecting ``optimal wordings'' based on our judgment, which might not reflect how participants understand these statements, we included various phrasings and let factor analysis reveal nuances in how readers interpret and respond to trust-related prompts (Section~\ref{exp1}). %This approach also ensures a broader and more representative measurement of trust dimensions.

Following the integrated framework of trust in visualizations~\cite{elhamdadi2023vistrust}, we also categorized items based on their association with \textit{cognitive} or \textit{affective} trust.
Cognitive trust is rooted in logical understanding and evidence, similar to the deliberate and analytical ``System 2'' way of thinking from the cognitive psychology literature~\cite{kahneman2011thinking}. In contrast, affective trust is driven by emotions and intuitions, reflecting the faster, instinctive ``System 1'' mode of thinking~\cite{kahneman2011thinking}. These two categories are orthogonal, as a reader may distrust a visualization due to negative emotional reactions to its topic (e.g., triggered by prior beliefs), even when there is no evidence suggesting that the visualization is designed to deceive or that the data is biased. We also clustered statements by topic, such as clarity, accuracy, and benevolence, drawing inspiration from the trust-related factors in Bhattacherjee et al.~\cite{bhattacherjee2002individual} and Mayer et al.~\cite{mayer1995integrative}. We emphasize that this clustering is preliminary and primarily serves to enhance the clarity of presentation at this stage. In the following sections, we describe a pilot study designed to identify visualization stimuli for testing this inventory. The data collected from this study will allow us to conduct a factor analysis to uncover true statistical clusters behind trust.

The aforementioned compilation resulted in 92 questions on trust in visualizations. Since personality psychology research has shown that a person's tendency to trust is a stable personality trait, we further included the 21 questions from Evans et al. measuring general trust and trustworthiness~\cite{evans2008survey}. These items could provide baseline information on participants' general trust dispositions, allowing a more contextualized understanding of their responses to visualization-related trust measures.

\section{Pilot Choosing Visualizations}
To select the most representative and diagnostic statements for measuring trust in visualizations, we first needed to collect data on a diverse set of visualizations using all inventory items. This would allow us to identify robust and informative items that effectively differentiate between visualizations with varying levels of trustworthiness and discard non-diagnostic ones. To achieve this, our visualization stimuli must encompass a broad range of trustworthiness across various dimensions. %all dimensions outlined in our inventory (see Tables \ref{trustQuestions1} and \ref{trustQuestions2}). 
Since a validated inventory did not yet exist, we conducted this pilot study by presenting participants with visualizations and asking them to report their trust in the visualization's creator, data, and design on straightforward statements that we construct heuristically.

% To select the most representative and diagnostic question items for measuring trust in visualizations, we need to test all the inventory items and collect data across a set visualizations. To make sure that (1) robust and informative items stand out enough to differentiate between visualizations that vary in trustworthiness, and (2) that non-diagnostic items can be effectively discarded, we need this set of stimuli visualizations to cover a wide range of trustworthiness across all dimensions we identified in our large inventory (see Tables \ref{trustQuestions1} and \ref{trustQuestions2}). Since we do not have a reliable and valid inventory to work with yet, we conduct this pilot study by showing participants visualizations and ask them to report their trust in the visualization's creator, data, and design \textit{heuristically.}

\vspace{1mm}
% We conduct a pilot study to identify a diverse set of data visualizations with varying levels of trust based on heuristics.
% later test our initial collection of trust inventory items.
% we conducted a pilot study to identify a diverse set of data visualizations to later test our initial collection of trust inventory items.

\noindent \textbf{Participants:} We recruited 60 participants ($M_{\textit{age}}$ = 30.3, $SD_{\textit{age}}$ = 11.23) via Prolific.com~\cite{palan2018prolific} and compensated them at a rate of \$12.01 per hour. 
We filtered for participants who were fluent in English, over 18 years old, and had an approval rate between 95-100\%. 
\vspace{1mm}
% After filtering responses to remove low-effort participants who completed the study in under 10 minutes (the minimal time required to provide quality response based on experimenter piloting), we were left with 60 participants (X women, X men, X non-binary or indicated to prefer to self-describe, $M_{\textit{age}}$ = 30.3, $SD_{\textit{age}}$ = 11.23).
% \cx{need to update and fill in demo stats}

\noindent \textbf{Stimuli, Design and Procedure:}
Since an adequate measure of trust for visualizations did not yet exist, selecting a set of visualizations with varying levels of trustworthiness posed a challenge. To identify appropriate visual stimuli for this pilot, we first curated a collection of \re{static} visualizations from online sources, including the MASSVIS Dataset~\cite{borkin2015beyond}, consisting of visualizations that represented varying degrees of trustworthiness based on our expert judgment. For example, we considered factors such as the source of the visualization, the visualization techniques used, and the credibility of the conveyed data. The resulting corpus consists of 40 visualizations, including five line charts, four pie charts, three bar charts, five infographics, nine line charts, six pie charts, and eight geo-spatial visualizations. These visualizations also span a wide range of subjects, covering topics such as sports, public health, finance, and physical science. Details can be found in Supplemental Materials.

We emphasize that these selections reflected our expert judgment and served only as a starting point. The ultimate goal was to collect participants' opinions on these visualizations and select a subset exhibiting diverse trustworthiness profiles according to actual reader perceptions rather than our assumptions. To this end, we asked participants to rate the trustworthiness of the creator(s), data, and design of these visualizations, following the trust framework outlined by Elhamdadi et al. ~\cite{elhamdadi2023vistrust}.
Specifically, they were given the statement ``I trust the [\textit{creator/data/design}] of this visualization'' and a scale ranging from 1 to 6 (1 = Strongly Disagree, 6 = Strongly Agree). For each dimension, participants also had the option to select ``Not Applicable.'' We emphasize that the purpose of this assessment was not to evaluate the inherent trustworthiness of the visualizations, but to identify those with varying levels of trustworthiness driven by different implicit trust factors that subsequent experiments can diagnose.

%  
% three bar charts, five line charts, four pie charts, two scatterplots, two geospatial visualizations, three infographics, and one bubble chart. (original 20)

% We separated the visualizations into two sets of 20 to keep the experiment length manageable. 

% (1) cover a diverse range of chart types, (2) cover a wide range of trustworthiness levels, and (3) elicit (dis)trust due to different reasons. To this end, the authors searched for charts online, incorporating selections from the MASSVIS Dataset\cite{borkin2015beyond}, and curated a collection that met the outlined criteria based on our judgment. 

% \begin{figure}
%     \centering
%     \includegraphics[width = 0.6\columnwidth]{Figure/line8.png}
%     \caption{The creator of this chart is the Wall Street Journal (see bottom right of the chart); the data is the numbers represented by the chart (e.g., one Groupon share was valued wat \$25 on Feb 8); the design captures all choices the creator made in crafting the chart, such as the use of a line chart, the choice of green for the line, and the addition of gridlines. Figure taken from~\cite{ovideWSJ}.
%     }
%     \label{fig:line8}
% \end{figure}

After consenting to the study, participants read a short introduction explaining our definitions of a visualization's creator, data, and design. 
% Figure~\ref{fig:line8} shows an example. 
They then read through 20 visualizations presented in random order and rated their trust in their design, data, and creator. %  of each. 
Next, participants ranked these visualizations by the extent to which they trusted them.
At the end of the survey, they reported demographic information 
and were given a quick debrief on the goals of the survey.
\vspace{0.5mm}

% ead 20 visualizations in random order and rated on a scale from ... to ... on how much they trusted the visualization design, the data visualized, and the creators of the visualization.
% \cx{insert quotes of the original text of the question}

\begin{figure}[t]
 \centering 
 \includegraphics[height = 0.5\columnwidth]{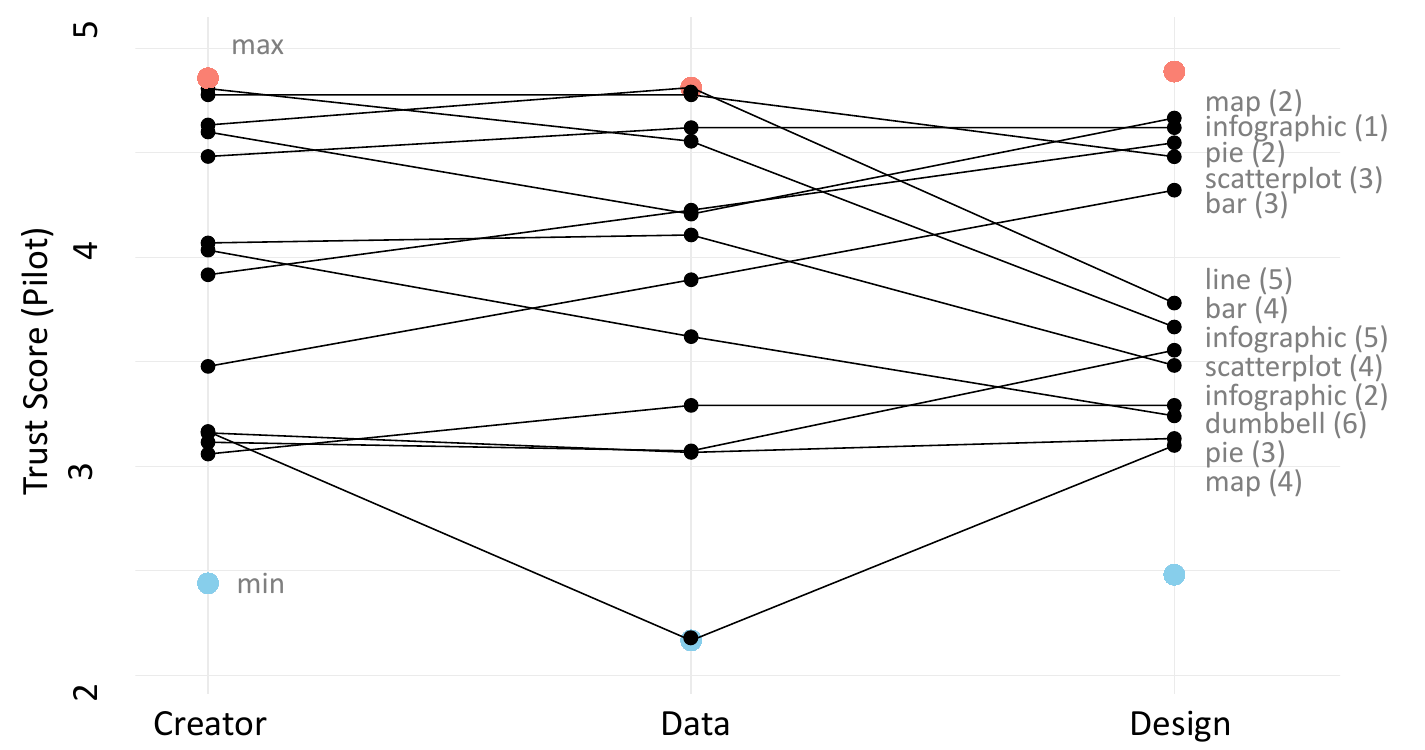}
 \caption{Mean trust ratings for the creator, data, and design dimensions across the 13 visualizations selected from the pilot for the main experiment. Red and blue points indicate the minimum and maximum mean trust ratings observed for any visualization within each dimension. The selected visualizations encompass a diverse range of trust ratings, with some consistently high or low across all dimensions, while others exhibit mixed trust profiles---high in some dimensions and low in others.}
 \label{pilot_vis}
\end{figure}

\noindent \textbf{Results:}
Since our goal was to curate a set of visualization stimuli with varying degrees of trustworthiness across different trust dimensions, we analyzed the distribution of visualizations along the preliminary trust dimensions. Specifically, we performed k-means clustering on the 40 visualizations separately for each of the three dimensions: creator, data, and design. 
Silhouette scores~\cite{shahapure2020cluster} indicate that the optimal clustering solution for each dimension involves two clusters. Therefore, we divided the visualizations into two subsets per dimension to capture both high and low trust ratings.
To ensure rich stimuli for downstream experiments, we included not only visualizations that were consistently (un)trustworthy across all dimensions but also those that demonstrated significant variation in trustworthiness between dimensions.
%Our goal was to select visualizations that collectively covered a broad range of trust ratings within each dimension while also exhibiting diverse patterns of trust across dimensions. To ensure rich stimuli for downstream experiments, we included not only visualizations that were consistently (un)trustworthy across all dimensions but also those that demonstrated significant variation in trustworthiness between dimensions.
To do so, we first identified visualizations within the top one-third and bottom one-third of trust ratings for each dimension. 
Within these subsets, we selected visualizations that showed the largest and smallest differences between all pairs of dimensions (design-data, design-creator, data-creator). After repeating this selection process for all three dimensions and removing duplicates, we obtained a set of 13 visualizations \re{spanning eight chart types}. Figure~\ref{pilot_vis} illustrates the trust ratings of the selected visualizations, confirming that they collectively represent diverse trustworthiness profiles. %In addition, these visualizations encompass eight distinct chart types, providing a varied typological sample that enriches our subsequent trust data collection.

% and their relative relation to each dimension. 
% Detailed views of the visualizations can be found in the supplementary materials. \cx{let's not forget to include this}

%%%%%%%%%%%%%%%%%%%%%%%%%%%%%%%%%%%
\section{Unpacking Trust in Visualizations via Exploratory Factor Analysis}
\label{exp1}
We administered the full set of 113 trust items to participants, using the 13 visualizations selected from the pilot study as stimuli. We then performed a post-hoc factor analysis to identify the underlying dimensions of trust within this set. Based on these results, we developed a significantly more compact inventory, retaining representative and diagnostic items.

% , which helps us understand which questions most diagnostically measure a viewer's trust in a visualization
% In an kylie{online?} study, we had participants go through the entire inventory of trust-related questions, which 
% This allows us to determine which trust items most diagnostically measure trust in visualizations.

\subsection{Study Design and Procedure}
\label{exp1design}
Given the potential for participant fatigue from the extensive set of trust-related items, we conducted a between-subjects experiment where each participant was randomly assigned to work closely with one of the 13 visualizations chosen from the pilot experiment.
Upon consenting to the study, participants engaged in a warm-up phase designed to prompt them to start thinking about \textit{trust} in the context of the assigned visualization.
During this phase, participants read the visualization and provided initial impressions. Specifically, they rated how much they trusted the visualization on a Likert scale. Since this overlaps with an existing item in the inventory, responses were excluded from the subsequent factor analysis to avoid redundancy.
To foster familiarity, participants were instructed to distill their main takeaways from the chart. In addition, they were asked to speculate on the creator and the intended audience of the visualization and comment on the data source.
% They speculated the visualization's key message and commented on the data source (i.e., where do the data for this chart come from? If it is not apparent, type `I don't know.'). They were asked about the creator of the chart and for whom they think the visualization is designed. 

%The purpose of this warm-up phase was to encourage participants to start thinking about \textit{trust} in the context of the given visualization before interacting with the inventory. 

% rating how much they trusted the visualization. This warm up overlaps with existing items in the inventory, and therefore participants' responses are not included in the upcoming factor analysis to avoid redundancy issues. Its purpose was to encourage people to start thinking about trust as they answer the remaining questions. 
% They then engaged with the visualization by speculating 

Next, they completed all items in the initial inventory, excluding those related to individual characteristics. They rated how much they agreed with each statement on a scale from 1 (Strongly Disagree) to 6 (Strongly Agree). 
To mitigate fatigue, we divided the questions into four blocks and encouraged participants to take a break at any time. 
At the end of each block, participants were invited to provide feedback on the items they had rated, particularly noting if any statements were unclear or confusing.
% At the end of each block, we asked participants to report whether they had any comments about the items they rated, in case any item was confusing. % or perceived not applicable to the visualization they read.
% We also included a question from the BeauVis~\cite{he2022beauvis} inventory to evaluate the perceived aesthetic pleasure people derived from the visualization \kylie{correct this if I am wrong haha}.
% The original inventory included five items, but since He et al. \cite{he2022beauvis} demonstrated that they are all closely associated with the same factor, we opt to include the Beauvis item with the highest factor loading: ``to what extend do you agree that this visual representation is likable." (1 = strongly disagree, 7 = strongly agree)
% We requested the participants to respond based only on the design of the visualization. 
% To acquire more specific reactions, we also asked participants to identify from a list of design features what they liked and disliked about the visualization, including the chart type, the color choices, the amount of text, the granularity of the data shown, the annotations, the legend, and the amount of data shown. 
% This information will be used later on to identify potential relationships between chart design and trust levels.
% \cx{don't forget to do this}
 %chart understanding by asking four visual literacy questions. 
Finally, participants completed the individual characteristic items measuring general trust and trustworthiness from the inventory (\#93–\#113)~\cite{evans2008survey}. % , which were included towards the end, separate from the visualizations, because they are more about measuring personality traits. 
The survey ended with demographic questions and a debrief summarizing our study goals. %  of our experiment.
% Detailed information of the experiment can be found in \cx{OSF link}

\subsection{Participants}
We recruited 270 participants via Prolific.com~\cite{palan2018prolific} and compensated them at a rate of \$12 per hour. On average, 20.8 participants were assigned to view and respond to questions about each visualization. We applied the same exclusion criteria as in the pilot study, with an additional filtering metric based on responses to the reverse-worded items in the inventory. We expected reasonable participants to report opposite attitudes on these reverse items. For example, if a participant disagreed with the statement ``I like this visualization'' (\#17), they should agree with ``I dislike this visualization'' (\#18). Out of the six pairs of reverse items, we counted the number of pairs for which a participant provided responses that were not directionally opposite (e.g., Strongly Disagree for \#17 and Disagree for \#18). Participants with three or more such inconsistencies were excluded from the analysis. After applying the exclusion criteria, we were left with 226 participants (82 women, 139 men, 5 non-binary or preferred to self-describe; $M_{\textit{age}}$ = 31.54, $SD_{\textit{age}}$ = 11.16).

% For example, if they ``disagree'' with the statement that they ``like this visualization'' (\#17), they should `agree' with `I dislike this visualization' (\#18). Out of the six pairs of reverse items, we counted the number of pairs for which a participant provided responses that were NOT semantically opposite (e.g., `strongly disagree' for \#17 and `disagree' for \#18), and excluded everyone with three or more counts. 

%%%%%%%%%%%%%%%%%%%%%%%%%%%%%%%%%%%%%%%%%%%%%%%%%%%%%%%%%%%%%%%%%%%%%%%%
%%%%%%%%%%%%%%%%%%%%%%%%%%%%%%%%%%%%%%%%%%%%%%%%%%%%%%%%%%%%%%%%%%%%%%%%
%%%%%%%%%%%%%%%%%%%%%%%%%%%%%%%%%%%%%%%%%%%%%%%%%%%%%%%%%%%%%%%%%%%%%%%%
\subsection{Results: Exploratory Factor Analysis}
We now present the results of an Exploratory Factor Analysis (EFA). The purpose of the EFA is twofold: it  aims to uncover the main ``themes'' behind a visualization reader's notion of trust, while also identifying a significantly reduced set of questions that represent each dimension. To ensure that the collected data is suitable for EFA, we first conducted Bartlett’s test of sphericity~\cite{bartlett1954note} and the Kaiser-Meyer-Olkin (KMO) test~\cite{kaiser1974index}. Bartlett’s test of sphericity confirmed sufficient correlations among items, with $p < 0.001$, indicating that the data is appropriate for factor analysis. The KMO test produced an overall measure of 0.94, suggesting excellent sampling adequacy and strong suitability for EFA.

% We conduct an EFA to reduce the number of items in our trust inventory to include only orthogonal items that cover the most distinct dimensions of trust. This analysis also captured the main `themes' of trust participants care about when evaluating a visualization, allowing us to create a compact trust inventory that can be easily applied in visualization assessment. 
We performed the EFA in R using the \verb|Psych| R package \cite{revelle2015package}. We provide the data and analysis script in Supplemental Materials. We first generated a correlation matrix considering all 113 items from our initial inventory. We observed strong correlations among multiple items. For example, the item ``I find this visualization easy to use'' and the item ``Learning to read this visualization was easy for me'' were highly correlated ($r = 0.81$), presumably because both capture visualization usability. 
Conversely, some items showed minimal correlation. For example, the item ``The data source of this visualization has a good reputation'' and the item  ``Learning to read this visualization was easy for me'' were weakly correlated ($r = 0.16$). These findings confirm that trust in visualizations is a multifaceted construct.

\begin{figure}[t]
  \centering
  \includegraphics[width = \columnwidth]{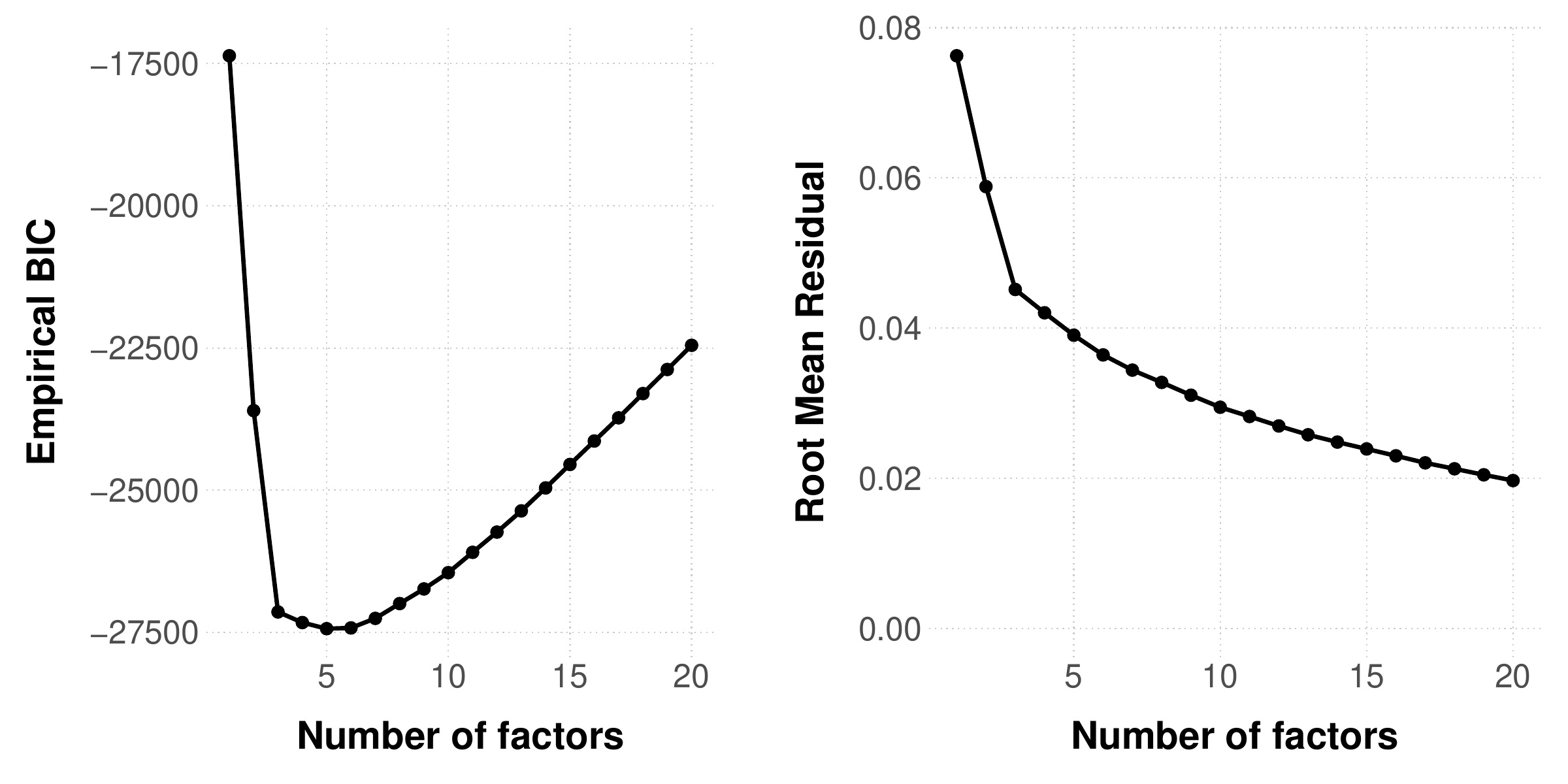}
  \caption{Comparison of factor models based on empirical BIC and root mean residual. Both metrics exhibit an ``elbow'' at the three-factor model, suggesting that it provides a favorable balance between model fit and parsimony.}
  \label{fig:factor}
\end{figure}

We compared the empirical Bayesian Information Criterion (BIC), root mean residual, and model complexity across factor models ranging from 1 to 20 factors, as shown in Figure~\ref{fig:factor}. We selected an oblique rotation method (oblimin) to allow for the possibility that different facets of trust may not be strictly independent, but rather share underlying conceptual relationships. The BIC, which balances model fit against complexity to prevent overfitting, dropped significantly when the model included three or more factors, showing a clear ``elbow'' at this point. While model complexity---the extent to which items load onto multiple factors---increased steadily with the addition of more factors till a seven-factor model, the root mean residual, which measures the average difference between observed and predicted correlations, decreased. Notably, the rate of root mean residual reduction diminished substantially after three factors, indicating diminishing returns for additional complexity. Based on these comparisons, we selected the three-factor model as the most parsimonious and informative representation of the 113 trust items.

The resulting model has a mean item complexity of 1.2, indicating that each item primarily loads onto one factor on average, which reflects good clarity in factor separation. The RMSR is 0.05 after adjusting for degrees of freedom, which suggests a good fit between the observed data and the factor model. Three items failed to load onto any factors. 
These items are ``I am taking a chance interacting with this visualization,'' ``I believe that people seldom tell you the whole story,'' and ``The content of the visualization is predictable.'' 
\re{We provide detailed factor assignment and loadings for all items within the three-factor model in Supplemental Materials.}
Two authors iteratively reviewed the item clusters for each factor and distilled the overarching themes, which we describe below:

% \will{removed TLI index and BIC for now (see comments); probably need to add variance explained and eigenvalues for factors}
% The Tucker Lewis Index of factoring reliability stands at 0.79, and the BIC is -23487.76.

% We found that empirical BIC dropped significantly once the model considered \textit{three} or more factors. Complexity generally increases with additional factors and root mean residuals decrease with more factors, but the rate of decrease drops after \textit{three} factors. Based on these model comparisons, we adopt the factor model with \textit{three} factors. 
% as the model to most optimally and efficiently capture the 113 items in our trust inventory.

% We first examined how factors within each model correlate with each other. Ideally, to create the best taxonomy, a good set of factors should correlate with each other as little as possible. There was no significant difference between the correlation between the model factors, with the highest correlation between factors in the model is 0.24 in all three models. 

\vspace{1mm}
\noindent \textbf{1. Trust in the Information:} This factor primarily encompasses items from the \textit{Trust in Creator} and \textit{Trust in Data} dimensions. This suggests that participants conflate trust in the data and the creator when assessing a visualization’s credibility. 72 items (63.7\%) load onto this factor, indicating its importance in shaping trust.
We evaluate the strength of association between an item and a factor with factor loadings~($\lambda$), which measure the correlation between an item and a factor.
Examples of items strongly associated with this factor are ``I trust this data'' ($\lambda = 0.93$), ``I believe that the creator(s) does not falsify data'' ($ \lambda = 0.92$), ``The data source of this visualization has a good reputation'' ($ \lambda = 0.72$), and ``The creator(s) is fair in the process of creating this visualization'' ($ \lambda = 0.84$). 
Examples of items weakly associated with this factor are ``I understand what this visualization is trying to tell me'' ($ \lambda = 0.18$) and ``I am taking a chance interacting with this visualization'' ($ \lambda = -0.22$). %, and ``I have a good word for everyone'' ($ \lambda = 0.02$). 

\vspace{1mm}
\noindent \textbf{2. Clarity, Usability, and Likability:} This factor primarily reflects the comprehensibility, usability, and likability of a visualization. A total of eighteen items (15.9\%) load onto this factor, including many measuring the \textit{Trust in Design} dimension. This suggests---perhaps unsurprisingly---that a visualization’s design plays a crucial role in how easily it can be understood, how effectively it can be used, and how much people like it.
Examples of items strongly associated with this factor are ``I find this visualization difficult to use'' ($ \lambda = -0.86$), ``[L]earning to read this visualization was easy for me'' ($ \lambda = 0.94$), and ``I dislike this visualization'' ($ \lambda = -0.57$). 
Examples of items weakly associated with this factor are ``The data source of this visualization is clear'' ($ \lambda = 0.01$) and ``I am filled with doubts about this'' ($ \lambda = 0.00$). 

Interestingly, this factor has a moderate correlation with the first factor ($r = 0.56$). Such correlation between factors is permitted under the oblique rotation method we employed. This relationship may indicate that when viewers find a visualization easy to understand and use, they may be more inclined to trust the underlying data and creator. Alternatively, when viewers already trust the source and data, they might perceive the visualization as more usable and clear. Despite this correlation, we emphasize that factors 1 and 2 remain distinct dimensions, as evidenced by the minimal cross-loading of items implied by the low mean item complexity (1.2) in our model.

\vspace{1mm}
\noindent \textbf{3. Individual Characteristics:} This factor primarily captures participants’ general tendency to trust and their perceived self-trustworthiness. A total of twenty items (17.7\%) load onto this factor.
% Items that are loaded into this factor mostly measure of the participants' own tendency to trust and perceived self-trustworthiness. Twenty items (17.7\%) loaded into this factor. 
Examples of items strongly associated with this factor are ``Respect others'' ($ \lambda = 0.64$) and ``Can get along with most people' ($ \lambda = 0.63$). 
Examples of items weakly associated with this factor are ``The creator(s) has the skills and expertise to make the most informative version of this visualization'' ($ \lambda = -0.01$) and ``The data in this visualization is unbiased and trustworthy'' ($ \lambda = 0.00$).

% We then observe the relationship between trust %, measured by the diagnostic items we extracted from the factor analysis, 
% and a few other measures captured in our experiment, including visual literacy and aesthetic preferences.

\section{Selecting Representative Survey Items}

In the previous section, we uncovered latent factors underlying what readers mean when they say they ``trust'' a visualization with EFA. The reduction of 113 items into three distinct factors suggests that many items in fact measure the same dimensions. In this section, we focus on reducing the number of items to create a more concise inventory. Ideally, selected items should both strongly represent their respective factor and effectively distinguish between visualizations of varying trustworthiness.
We begin by introducing the concept of \textit{item discriminability}, which reflects how well an item distinguishes between trustworthy and untrustworthy visualizations. We then select representative items for each dimension based on their factor loadings and item discriminability.

\subsection{Item Discriminability}
\label{discriminability} 
Since the 13 visualization stimuli selected from the pilot represent a broad range of reader-perceived trustworthiness across creator, data, and design, a ``good'' question should accordingly capture the variation among stimuli.
To quantify the ability of an item to reflect this variation, we computed a \textit{discriminability score} for each item. Specifically, we first calculated the average participant rating for each item on every visualization. Next, for each item, we computed the difference in average ratings across all 78 unique pairs of the 13 visualizations. We then averaged these 78 pairwise differences to obtain the item's overall discriminability score, which reflects how effectively the item distinguishes between visualizations. \re{The discriminability scores for all items are sorted and visualized in Supplemental Materials.}
%The discriminability scores are sorted and presented in Figure \ref{fig:discriminability}. (say its in supp. moved to supp as well)

% We first computed the discriminability of the trust inventory items from Tables \ref{trustQuestions1} and \ref{trustQuestions2} to identify diagnostic and representative items to form a compact list of questions for measuring trust in visualization.
% Since the 13 visualizations were chosen to maximize variation on reader-perceived trustworthiness in creator, data, and design (amongst the 40 visualizations we piloted), we expected a relatively large difference between at least some pairs of visualizations on any item.
% For each item, we computed the average rating of that item for each visualization. 
% Then, for each pair of visualizations (13 visualizations, 78 unique pairs), we computed the difference between their average ratings on this item. 
% We averaged the 78 pair-wise differences to generate the a \textbf{discriminability} score, which are sorted and shown in Figure \ref{fig:discriminability}. 

% \begin{figure}[!t]
%   \centering
%   \includegraphics[width = 0.86\columnwidth]{Figure/item_scores.pdf}
%   \vspace{-5pt}
%   \caption{Mean discriminability scores of visualization-related trust items. Error bars represent 95\% confidence intervals. Items selected for the compact inventory are highlighted in red. Items are ranked by mean discriminability, which quantifies their ability to differentiate between trustworthy and untrustworthy visualizations.}
%   \label{fig:discriminability}
% \end{figure}

Items with higher discriminability scores more effectively differentiate between visualizations. For example, ``The data source of this visualization is clear'' and ``I find this visualization difficult to use'' exhibit high discriminability. Such items are typically concrete and straightforward. As a result, participants’ responses are more likely to be grounded in direct observations or measured extrapolations from the information in the chart.
In contrast, items with low discriminability fail to distinguish between visualizations, as participants provide similar ratings across different stimuli. These items often tend to require participants to make inferences or predictions in hypothetical scenarios. For example, ``I am taking a chance if I were to make a decision with this visualization'' and ``I believe that the creators will update this visualization when new information is available'' have low discriminability. Their speculative nature might have made them more challenging for participants to evaluate.
Overall, we prioritize including high-discriminability items to represent the trust dimensions.

% Pair-wise comparison for all 13. 
% Items ranked from most to least differentiable. 
% Also looked at design vs. data, design vs. creator, data vs. creator, like Andrew suggested. 

\subsection{Selecting Representative Survey Items} 
We began by selecting items with the highest factor loadings from each of the three factors. Items with the highest loadings are ones that most strongly correlate with and represent their respective trust dimension. Specifically, for each factor, we retained two positively worded and two negatively worded items with the highest factor loadings. These high-loading items are theoretically the most predictive of their respective factors.
For example, for Factor 1 on \textit{Trust in the Information}, the four corresponding items are 
(1) I trust this data (+); 
(2) I believe that the creator(s) does not falsify data (+); 
(3) I am skeptical about the information presented in this visualization (-); and 
(4) The creator(s) believes that it is okay to lie with visualizations (-). 

Including multiple items associated with the same factor allows administrators of our inventory to assess response reliability and identify participants who respond randomly. 
For example, a participant who provides high ratings for both Item (1) and Item (3) would be flagged for inconsistency, indicating potentially unreliable data.
Additionally, including both positive and negative items helps mitigate response bias~\cite{mcgrath2010evidence}, where participants might indiscriminately provide uniformly high or low ratings, such as by moving the slider up to high values for all questions. %By comparing responses to items with opposing valence, administrators can detect and filter participants exhibiting such patterns.
To ensure that the final survey is diagnostic, we also considered the discriminability of each item. Using the previously computed discriminability ratings, we retained the most discriminative positive and negative items for each factor. \re{For Factor 1, the selected items were \#43 and \#47; for Factor 2, the selected items were \#4 and \#14.}
%For Factor 1, the final selected items were Item~(1) and Item~(3). 

% We also care about the discriminability of each item to ensure that the final survey is highly diagnostic and compact. 
% Based on our discriminability ratings calculated from Section~\ref{discriminability}, among the four items associated with each factor we selected, we choose to retain the positive and negative items with the higher discriminability, which happens to be (1) and (3). 
% Including both positive and negative items will allow users of our inventory to mitigate response bias, such that participants who are indiscriminately responding in one direction (i.e., moving the slider up to high values for all questions) can be identified and filtered out. 
% both positive and negative items will allow users of our inventory to measure response reliability and mitigate 'positive bias' \cite{mezulis2004there}.
% Comparing the consistency between how participants responded on these four items will allow researchers to filter out participants who responded randomly.

For Factor 3 on individual characteristics, however, we did not have discriminability data because the questions measure participant-dependent traits rather than chart-dependent properties. Instead, we retained the two positive and two negative items with the highest factor loadings. 
% They serve as a baseline measure of an individual's tendency to trust and their trustworthiness.
% We elect to retain the two positive and two negative items with the highest loading.
These items cover both tendencies to trust and self-perceived trustworthiness of a person. 
This resulted in an inventory with four visualization-related items  (\#4, \#14, \#43, \#47) and four individual characteristics items (\#95, \#104, \#107, \#110).

\section{Validating the Compact Trust Survey}
To validate the compact inventory, we \re{first evaluated its \textbf{content validity}---the extent to which it accurately and comprehensively measures the target construct. Our initial item compilation drew from diverse sources to ensure broad coverage of trust-related dimensions, and our factor analysis identified two distinct factors underlying trust in visualizations and another factor on general trust disposition. Our final inventory covers all three dimensions. The four visualization-related items capture usability, comprehensibility, and credibility---constructs well-established in the visualization trust literature~\cite{elhamdadi2023vistrust, pandey2023you}. The four general trust disposition items are drawn from a validated inventory~\cite{evans2008survey} and all load strongly onto a single factor, reflecting interpersonal trust behaviors such as openness to contact with others and ``having a good word for everyone.'' Together, these characteristics establish strong content validity.}

\re{The remainder of this section is dedicated to assessing the} \textbf{criterion validity} of our inventory---the extent to which it predicts theoretically related behavior~\cite{james1973criterion}. We focused on criterion validity because trust plays a crucial role in shaping real-world decision-making and interactions~\cite{punyatoya2019effects, mayr2019trust}. Specifically, we compared the trust measurements collected with our inventory against \re{behavioral} outcomes from an adapted investment trust game.

The classic trust game involves two parties: a trustor and a trustee. The trustor is given a sum of money and can choose to send a portion to the trustee (a person or a program)~\cite{ANDERHUB2002197, gaba2023my}. The sent amount is tripled, and the trustee may return some or none of the money to the trustor~\cite{BERG1995122, zheng2002trust}. The trustor’s willingness to send money, knowing they risk losing it, is considered a behavioral measure of trust~\cite{brulhart2012does}. 
However, reading a visualization fundamentally differs from interacting with another person or agent: visualization consumption typically involves a one-directional information flow from the visualization to the reader, which lacks the multi-round, interactive dynamics that characterize traditional trust games. While trustees in classic trust games can directly diminish the trustor's welfare, visualizations' potential for harm lies in their capacity to propagate misinformation that may lead viewers to make self-defeating decisions.

\begin{figure*}[!t]
  \centering
  \includegraphics[width = \textwidth]{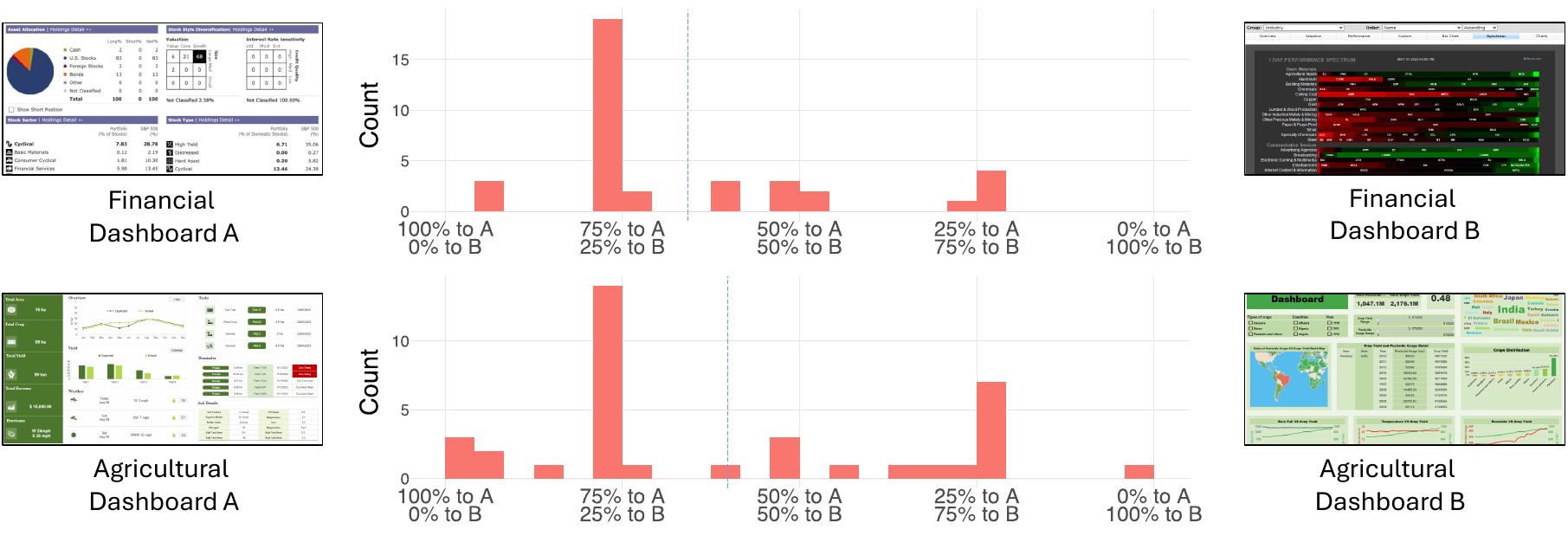}
  \caption{Distributions of resource allocations in the trust game. In both investment contexts, Dashboard A receives significantly more resources, indicating higher behavioral trust among participants.}
  \label{fig:allocation}
\end{figure*}

% The classic trust game typically involves two roles: a trustor and a trustee.  However, viewing a visualization fundamentally differs from interacting with a human or an agentic system. Unlike in classic trust games, where the trustee can harm the trustor through direct reaction, a visualization cannot harm the trustor the same way. Fundamentally, a typical visualization viewing experience only entails a one-directional information flow from the visualization to the reader, lacking the multi-round, inter-party dynamics that inform traditional trust games. 

Since a core function of visualizations is to support decision-making, we framed our study within an investment scenario. To strengthen the ecological validity of our experiment, we introduced real-world stakes, a practice shown to enhance the reliability of trust assessments~\cite{alos2019trust}. %Specifically, participants were asked to make investment decisions based on the presented dashboards, ensuring their choices carried meaningful implications. 
To mitigate response consistency bias, where participants might align their trust ratings with their prior decisions to maintain cognitive consistency, we adopted a between-subjects design and conducted two separate experiments. %We now present these experiments in detail.

\re{In addition to establishing validity, we evaluated the inventory's reliability---the consistency of measurement. We calculated McDonald's omega ($\omega$)~\cite{Mcdonald1999TestTA}, an internal consistency measure generally preferred over Cronbach's alpha ($\alpha$) in contemporary psychometric practice~\cite{Hayes2020UseOR, Dunn2014FromAT}.}

\subsection{Trust Game}
\label{trust game}
To measure participants' behavioral trust in a visualization, we presented each participant with a hypothetical investment game where they viewed \re{a screenshot from each of} two data dashboards---each representing an advisor---and were tasked with allocating a fixed amount of financial or agricultural resources to each advisor based on the dashboard they were using.
\re{Consistent with our inventory development process, which utilized static visualizations throughout, we employed static screenshots of dashboards in this validation study. While the original dashboards may have contained interactive features, presenting them as static screenshots allowed us to maintain methodological consistency and exclude interactive features that may introduce additional facets of trust.}

% The baseline survey results suggest the Morningstar portfolio visualization to be more highly trusted.Now we investigate whether the trust game will yield similar results.
\vspace{1mm}

\noindent \textbf{Participants:} After conducting a pilot study with 20 participants, we performed a power analysis and ultimately recruited a total of 83 participants via Prolific.com~\cite{palan2018prolific} (48 men and 35 women; $M_{\textit{age}}$ = 32.96, $SD_{\textit{age}}$ = 12.03). 43 participants partook in the financial investment game, while 40 partook in the agricultural investment game.
The experiment took on average 5.93 minutes to complete, and participants were compensated at a rate of \$12 per hour.
\vspace{1mm}

\noindent \textbf{Procedure and Design:}
To validate our inventory across different contexts, we designed a between-subjects experiment with two investment scenarios---one familiar and one less familiar to participants. For each scenario, we sourced two different dashboards with distinct information and visual styles.
In the familiar context, dashboards displayed financial investment data, and participants were given a sum of \$10,000 to allocate. One dashboard was from Morningstar~\cite{morningStar} (Advisor A), while the other was from Finviz~\cite{finviz} (Advisor B). In the less familiar context, participants were tasked with distributing 100,000 crop seeds between the farmers. The dashboards presented agricultural data, and advisors were portrayed as farmers using agricultural analytics for decision-making. One dashboard was made by Kumar~\cite{kumar} (Advisor A), while the other was from InetSoft~\cite{inetsoft} (Advisor B).
Participants were randomly assigned to one context and presented with both dashboards in randomized order. We instructed participants that each advisor would use their respective dashboard for decision-making (financial or agricultural) and emphasized that allocations should be based on their impression of the dashboards.

Crucially, to strengthen ecological validity and ensure meaningful engagement, we informed participants that their Prolific compensation would increase proportionally based on the return on investment achieved by their advisors' decisions. This approach introduced real-world stakes, enhancing the reliability of our trust assessment~\cite{alos2019trust}.
Participants indicated how they would allocate their resources (money or seeds) between the two advisors using a continuous slider ranging from 0\% to Advisor A (100\% to Advisor B) at one extreme to 100\% to Advisor A (0\% to Advisor B) at the other. To ensure response validity, we included an attention check immediately after the allocation task that required participants to indicate which advisor they had invested in more heavily. Towards the end of the survey, we administered the general trust questions from our inventory and collected demographic information. All participants received a \$1 bonus payment regardless of their allocation decisions, and were debriefed about the experimental design.
\vspace{1mm}

\begin{figure*}[!t]
  \centering
  \includegraphics[width = \textwidth]{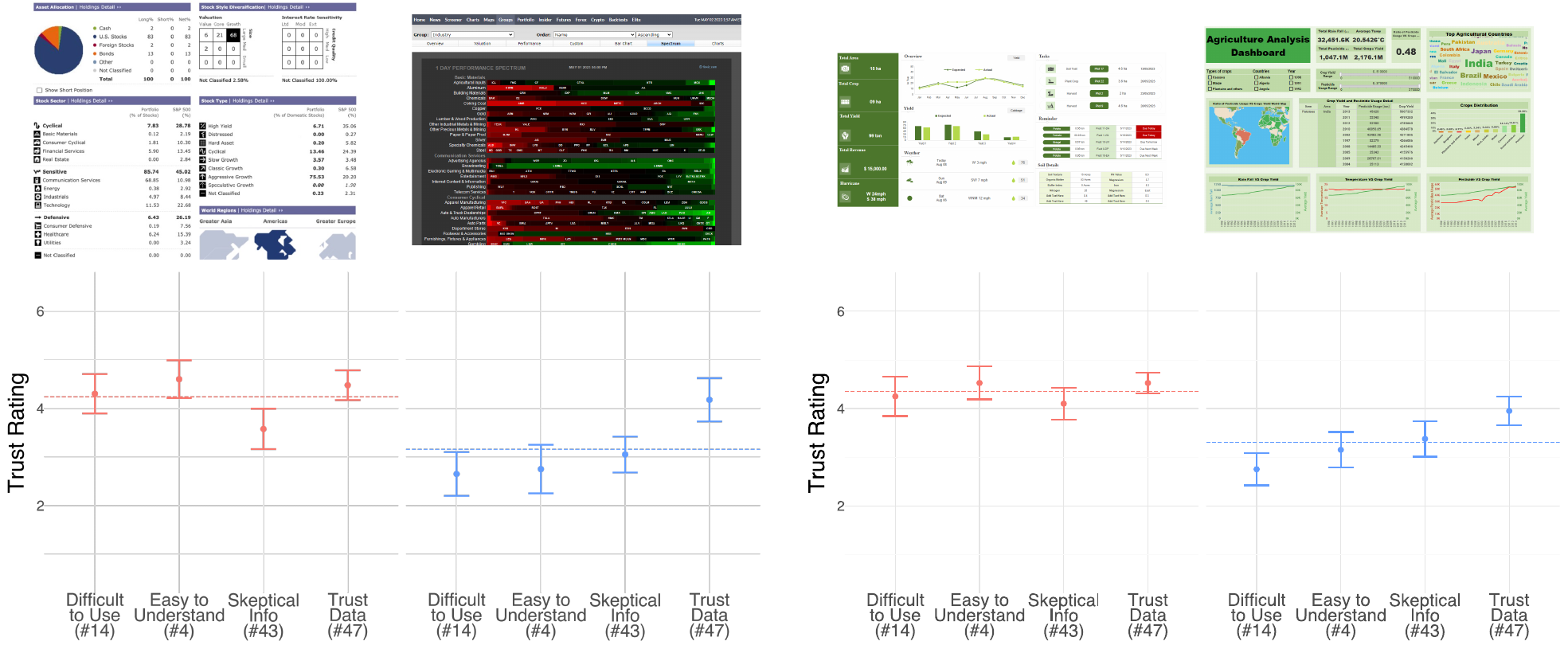}
  \caption{Viewer trust in visualizations as measured by the four visualization-related items in our inventory. Scores on negative items (\#14 and \#43) are reversed for clarity, ensuring that higher ratings consistently indicate greater trust. Dashed lines represent mean ratings. In both investment contexts, Dashboard A receives more resource allocations, supporting the criterion validity of our inventory in capturing behavioral trust.}
% for each visualization. }
  \label{fig:trust ratings}
\end{figure*}

\noindent \textbf{Results:} 
The amount of resources allocated to each advisor provides a behavioral measure of trust. After filtering out participants who failed the attention check, six participants from the financial investment game (13.95\%) and three participants from the agricultural investment game (7.5\%) were excluded from the analysis. Since each participant viewed the two dashboards in random order, we first compared participants' allocation decisions for the two orders via a two-sample t-test. In both investment contexts, we found that dashboard presentation order did not significantly affect participants' allocation decisions.

Figure~\ref{fig:allocation} shows the distributions of resource allocation in the two investment games.
In both investment games, participants allocated significantly more resources to Advisor A (financial: $t(36)=4.758$, $p<0.001$; agricultural: $t(36)=2.235$, $p=0.032$). On average, participants allocated 65.78\% of resources to Advisor A in the financial game (95\% CI = [59.06\%, 72.51\%]) and 60.16\% in the agricultural game (95\% CI = [50.94\%, 69.38\%]). We also examined whether participants' general trust tendencies predicted their allocation behavior by correlating scores on the general trust disposition items with their resource allocation decisions. This analysis revealed no significant correlations, likely because participants were splitting a fixed amount of resources between the two advisors. 
\vspace{1mm}

%%%%%%%%%%%%%%%%%%%%%%%%%%%%%%%%%%%%%%%%%%%%%%%%%%%%%%%%
% \subsection{Discussion}

\subsection{Trust Measurement using Inventory}
In this experiment, we collected trust measurements for the dashboard stimuli employed in the previous study using our inventory. We then analyzed whether dashboards receiving higher trust ratings on our inventory predicted greater resource allocation in the investment task. 

\vspace{1mm}

\noindent \textbf{Participants:}  
After conducting a pilot study with 20 participants, we performed a power analysis and ultimately recruited a total of 80 participants via Prolific.com~\cite{palan2018prolific} (44 men, 35 women, and one non-binary; $M_{\textit{age}}$ = 36.45, $SD_{\textit{age}}$ = 13.70). 40 participants partook in each investment game.
The experiment took on average 7.81 minutes to complete and participants were compensated at a rate of \$12 per hour.
\vspace{1mm}

\noindent \textbf{Procedure and Design:}
To ensure trust measurements collected with our inventory remained meaningfully grounded in a task context, we employed identical investment scenarios to those in our previous experiment. Participants were randomly assigned to either the financial or agricultural investment game and received the same instructions. After explaining the investment game, including the bonus rules, we first asked participants to evaluate the dashboards using our compact inventory within the context of the investment task. Subsequently, we presented participants with the same resource allocation questions as in the previous experiment. However, these allocation responses were excluded from our analysis to avoid potential response consistency biases---they were administered solely to maintain scenario coherence for participants. For behavioral measures, we instead utilized the allocation data collected in the previous experiment.

% In a within-subject setup, participants first viewed the two portfolios in random order and rated how much they agreed with the four inventory items associated with the visualization-specific factors in our compact survey (items \#4, 14, 43, 47). Then, they provided ratings to the remaining statements associated with the individual characteristics factor (items \#95, 104, 107, 110.) from Table \ref{trustQuestions2}). 
\vspace{1mm}

\noindent \textbf{Results:} 
Figure~\ref{fig:trust ratings} shows the mean trust ratings for each dashboard across all visualization-related questions, with scores on negatively worded items (\#14, \#43) flipped for easier comparison. Dashboard A was consistently rated as more trustworthy across all visualization-specific dimensions in both contexts (financial: $t(39) = 5.20$, $p < 0.001$; agricultural: $t(39) = 5.68$, $p < 0.001$). In the financial investment game, the average trust rating for Dashboard A was 4.24 ($SE = 0.249$), compared to 2.60 ($SE = 0.219$) for Dashboard B. In the agricultural context, Dashboard A received an average trust rating of 4.47 ($SE = 0.123$), while Dashboard B received 3.29 ($SE = 0.137$). As this was a within-subject experiment, participants' individual characteristics were naturally controlled for. Additionally, a two-sample t-test revealed no significant effects of dashboard presentation order.

These results from our compact trust inventory align with the behavioral data from the trust game in Section~\ref{trust game}, where participants allocated significantly more resources to Dashboard A in both investment contexts. The strong correspondence between participants' explicit trust ratings and their allocation behavior in a consequential investment task demonstrates the criterion validity of our inventory, confirming its ability to predict trust-related behaviors in real-world decision contexts.

\re{We also assessed the internal consistency of our inventory to establish reliability. Because our inventory comprises two distinct components---a required section measuring trust in visualizations and an optional section measuring general trust disposition---each assessing fundamentally different constructs, we calculated McDonald's omega ($\omega$) separately for the two to check if items within each component consistently measure their respective underlying constructs. Both the four questions measuring trust in visualizations ($\omega$ = 0.79) and the four questions measuring general trust disposition component ($\omega$ = 0.70) meet the commonly accepted threshold of 0.70 for reliability~\cite{devellis2021scale}, confirming that each part of our inventory reliably measures its intended construct.}

\section{Applying the Inventory: Three Scenarios}

The need to measure trust in visualizations arises in various research contexts. In this section, we illustrate how our trust inventory can be tailored to diverse research applications through three scenarios.

\subsection{Obtaining a Trust Measurement}

Akira, a visualization researcher, is evaluating a novel uncertainty visualization technique and needs a standardized way to measure how much users trust some visualization stimuli.

If Akira has a manageable number of visualization designs they want to test, they could employ a within-subjects design where all participants read every visualization. 
In this case, they could administer only the four visualization-specific items from our inventory, as individual differences in general trust disposition are naturally controlled for.
\re{To obtain a trust score for each visualization, Akira would average the ratings across these four items (after reverse-coding reverse-worded items). Higher average scores indicate greater trust in that visualization.}
% They could administer the entire inventory, which includes both the visualization-specific items (\#4, \#14, \#43, \#47) and the four individual characteristics items (\#95, \#104, \#107, \#110).

If, however, there are more visualizations than can be reasonably administered to every participant, Akira should use a between-subjects design where each participant reads only a subset of the visualizations. With a sufficiently large sample size, the individual differences in general trust tendencies would likely balance out.
% with a large enough sample of participants.
In this case, Akira could administer only the four visualization-specific items of the inventory. % , as random assignment would likely distribute participants with varying trust dispositions evenly across conditions. 
If Akira is restricted to a small sample size, they should administer the full inventory. 
We recommend using mixed-effects modeling with visualization as a fixed effect, participant as a random effect (if each participant sees multiple visualizations), and general trust tendencies (the individual characteristic items) as a participant-level covariate:

\begin{equation}
\text{Trust Rating} \sim \text{Visualization} + \text{General Trust} + (1|\text{Participant})
\end{equation}

\re{Here, Trust Rating is the averaged score from the four visualization-specific items (after reverse-coding) for each observation. General Trust is computed by averaging the four general trust disposition items for each participant, serving as a participant-level covariate that accounts for baseline differences in trusting tendency. In this mixed-effects model, Visualization enters as a categorical fixed effect, General Trust enters as a participant-level covariate, and the random intercept (1|Participant) accounts for baseline differences between participants. The model uses one visualization as a reference level (usually the first alphabetically or as specified by the researcher), and the fixed effect coefficients for the other visualizations indicate how much more or less trustworthy they are compared to this reference visualization, after adjusting for individual differences in general trust tendency. This approach allows Akira to partial out the effect of individual differences in trust disposition, providing a more precise estimate of the visualization-specific effects.}

% This approach allows Akira to partial out the effect of individual differences in trust disposition, providing a more precise estimate of the visualization-specific effects. The resulting model coefficients for each visualization represent their trustworthiness rankings after controlling for participant-level variance in general trust tendency.

%
\subsection{Using Trust as a Predictor Variable}
Bob is investigating how trust in data visualizations influences information recall in the context of personal health. 
He needs to understand which specific aspects of trust most strongly predict clinical decisions.
% In this case, Bob is treating trust as a predictor variable. 
% For researchers like Bob who aim to use trust as a predictor variable,

We recommend administering the full eight-item inventory in Bob's case, where he treats trust in visualizations as a predictor variable. 
Our factor analysis reveals that trust in visualizations encompasses multiple dimensions. 
By capturing all factors (visualization clarity, usability, data credibility, and information comprehensibility), % the information credibility factor and the usability/clarity factor, 
Bob can examine which dimensions of trust most strongly influence his dependent variable, information recall. 
Additionally, including the general trust disposition items enables researchers to control for individual differences, thereby isolating the unique contribution of visualization-specific trust to the outcome measures.

In practice, researchers can either use the individual items as separate predictors or create composite scores for each factor. 
For example, Bob might use a multiple regression model to predict decision quality from the two trust dimensions while controlling for general trust tendencies:

\begin{equation}
\begin{split}
\text{Information Recall} \sim & \beta_{1} \cdot \text{Trust in Information} + \\
& \beta_{2} \cdot \text{Usability} + \\
& \beta_{3} \cdot \text{General Trust} + \varepsilon
\end{split}
\end{equation}

\re{Bob would first compute average composite scores for each factor after appropriate reverse-coding. The regression coefficients ($\beta_1$, $\beta_2$, $\beta_3$) then reveal the relative importance of each trust dimension in predicting information recall, with significant positive coefficients indicating that higher trust on that dimension leads to better recall, for instance.}

\subsection{Adapting the Inventory for Specialized Trust Definitions}

Caitlin is studying how journalists evaluate visualizations for inclusion in news articles. 
She is specifically interested in measuring trust as it relates to journalists' willingness to republish and disseminate visualizations. While the trust inventory in its current form captures key aspects of readers' perceived visualization usability and clarity that would serve Caitlin's general research needs, she wants to further refine the measurement to fit her specific context.
Specifically, 
% Our operational definition of trust emerged through a bottom-up, data-driven process that identified broadly applicable dimensions. However, we recognize that trust manifests differently across contexts. In Caitlin's case, 
she seeks to emphasize trust as the confidence to use a visualization as a source fit for public dissemination. 

For researchers like Caitlin with specialized trust definitions, our complete dataset of 113 trust-related items with their associated factor loadings and discriminability scores provides a rich foundation. 
Instead of building an inventory from scratch, Caitlin can begin by identifying items from our inventory that align most closely with her definition. 
For example, to capture behavioral outcomes, she might select items such as ``I'd be comfortable sharing this visualization if someone asks me a question which can be answered by this visualization'' (item \#49). 
If she believes source credibility to be important for visualization republishing, she could include ``The creator(s) of this visualization is unbiased and trustworthy'' (item \#91).

When selecting items, Caitlin should prioritize those with high discriminability scores, as these items most effectively differentiate between visualizations with varying levels of trustworthiness. 
After selecting appropriate items, she should decide whether to supplement her research with new items. % to her battery of questions  some of the existing TrustVis items or supplement them with the newly identified questions. 
This decision hinges on how well the existing items align with her research goals and the extent to which the new items overlap with or diverge from the existing ones. 
\re{Once her custom item set is finalized, Caitlin would compute trust scores by averaging the selected items (after appropriate reverse-coding), yielding a single composite score tailored to her definition of publication-oriented trust.}
As a final step, Caitlin should validate her adapted inventory through an appropriate experiment.
For example, she might design an experiment correlating inventory scores with actual editorial decisions about visualization publication. 
This context-specific validation ensures her adapted inventory measures the particular type of trust relevant to her research questions.

\section{\re{Limitation}}

\re{While our study provides a compact, validated inventory for measuring trust in visualizations, several limitations warrant discussion.}

\noindent \re{\textbf{Item Coverage} During initial item compilation, we systematically drew from diverse sources across social sciences and visualization research, crossing multiple theoretical frameworks and generating numerous phrasings to capture trust nuances. Our final pool included 92 visualization-specific items and 21 general trust items. However, our item generation process involved author-created paraphrases and adaptations, which may introduce subjective biases in wording choices. Additionally, despite this comprehensive approach, some aspects of trust or particular phrasings may not be represented. Nevertheless, the diversity of our initial pool, the inclusion of items from validated inventories, and our data-driven factor analysis approach provide confidence that our inventory captures the primary dimensions of trust in visualizations.}

\noindent \re{\textbf{Behavioral Validation Scope} 
Our criterion validity assessment focused on resource allocation decisions across two distinct investment contexts (financial and agricultural). Our setup represents a robust test of trust measurement, as it involves real stakes and behavioral commitment rather than mere stated preferences.
While this validation establishes the inventory's predictive validity for allocation-type decisions, visualizations in practice support other diverse activities, such as knowledge acquisition and hypothesis generation. Additionally, our validation involved single decisions, whereas real-world visualization use often entails sequential decision-making. Nonetheless, our inventory's strong content validity---capturing fundamental dimensions of usability, comprehensibility, and credibility that underlie trust across contexts---suggests it should generalize well to these broader visualization use cases.}

\noindent \re{\textbf{Static Visualization Focus} Our inventory was developed and validated solely using static visualizations. Interactive visualizations may introduce additional trust dimensions beyond those captured by our inventory. Unlike static charts, interactive systems respond to user input, which may raise additional considerations about whether interactive behaviors function as expected. The act of interaction may also influence trust formation, potentially strengthening confidence through increased transparency or diminishing it when inconsistencies emerge. Future research could extend our methodology to interactive contexts by incorporating interaction-specific items during initial compilation and validating the resulting inventory's psychometric properties with interactive visualizations.}

\section{Conclusion}

% The present work compared 100+ question statements from existing inventories across social and computer sciences via a factor analysis to identify the most diagnostic and representative phrasing that researchers can use in their studies toward a unified trust survey. This goal addresses the issue that existing visualization trust measurements tend to use a variety of question phrasings, making it not only difficult to compare experimental results across multiple studies, as small perturbations in question phrasing can affect viewer responses (e.g., ~\cite{rasinski1989effect}, \cite{fan2010factors}).

The present work employed a data-driven approach to understand what visualization readers mean when they say they ``trust'' a visualization. By adapting and analyzing over 100 trust-related statements from existing inventories across social and computer sciences via factor analysis, we constructed an \re{eight-item} inventory\re{: four core items measuring
trust in visualizations and four optional items controlling for
individual differences in baseline trusting tendency.} These statements not only effectively represent each trust factor but also exhibit high item discrimination. \re{We further established the inventory’s internal consistency reliability using McDonald’s omega, confirmed its content validity by demonstrating alignment with theoretically-grounded trust dimensions, and validated its criterion validity through two trust games with real-world stakes. Our work} addresses a critical gap in visualization research, where varied trust measures have made it challenging to compare findings across studies---a significant problem given that even small variations in question phrasing can substantially affect viewer responses. 

%We identified survey questions with wording choices that most diagnostically represent user responses regarding trust in data visualizations.
Our findings indicate that people perceive a trustworthy visualization as one that presents credible information and is both comprehensible and usable. %Additionally, we found that general trust disposition influences how individuals assess visualization trustworthiness.
We see our quantification of the visualization-related factors behind viewer trust as an operational definition of trust, one that emerges from empirical data. 
Beyond enhancing our understanding of how viewers conceptualize trust, the data we collected around viewer trust also provide a valuable reference point for future research aiming to define and measure trust for specific purposes. 
For instance, when trust in visualizations is framed in terms of specific behaviors, researchers can leverage our EFA findings and item discriminability scores to select the statements most relevant to their research goals, enabling them to develop and validate customized trust inventories.
This adaptability allows our work to function both as an empirically grounded measurement tool and a foundation for context-specific trust research.

Why people (dis)trust visualizations potentially depends on a variety of factors. One such factor is visualization design. \re{Our inventory enables researchers to investigate how specific design choices---such as chart type, color scheme, and annotation style---affect trust. For instance, in conjunction with existing instruments like PREVis~\cite{previs} and BeauVis~\cite{he2022beauvis}, researchers can systematically examine whether and how aesthetic appeal and readability relate to trust, advancing our understanding of trust antecedents.} 
Additionally, our inventory's ability to account for individual trust dispositions makes it particularly valuable for studying how other reader characteristics (domain expertise, prior beliefs, data literacy) influence trust in data communication~\cite{peck2019data, reinecke2013predicting, xiong2022seeing}. \re{In particular, our inventory includes questions on visualization usability and comprehensibility. While visualization literacy likely affects how readers assess these factors, metacognitive biases such as the Dunning-Kruger effect~\cite{dunning} may lead them to misjudge their own level of comprehension, introducing discrepancies between perceived and actual understanding. Future research can leverage our inventory alongside visualization literacy measures (e.g., VLAT~\cite{lee2017vlat} and Mini-VLAT~\cite{minivlat}) to disentangle how skill level and actual comprehension jointly shape trust judgments.} With this validated instrument, we hope future research will provide deeper insights into how to cultivate appropriate trusting behavior in human-data interactions.

\section*{Acknowledgments}
This research is sponsored in part by the U.S. National Science Foundation through grants IIS-2320920 \& IIS-2311575.

\bibliographystyle{IEEEtran}
\bibliography{reference}

% \newpage

\begin{IEEEbiography}
[{\includegraphics[width=1in,height=1.25in,clip,keepaspectratio]
{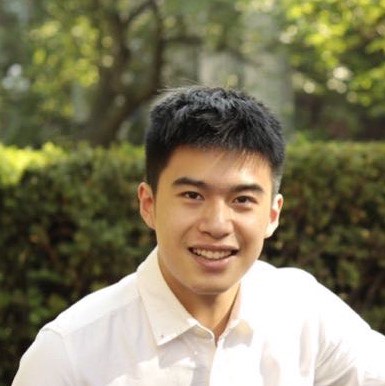}}]
{Huichen Will Wang} is a PhD student at the Paul G. Allen School of Computer Science \& Engineering at the University of Washington. %He holds a B.A. from Amherst College. 
His research focuses on developing and evaluating human-centric tools for data visualization and data science. He is also interested in graphical perception.
\end{IEEEbiography}

\vspace{-1.2cm}

\begin{IEEEbiography}
[{\includegraphics[width=1in,height=1.25in,clip,keepaspectratio]
{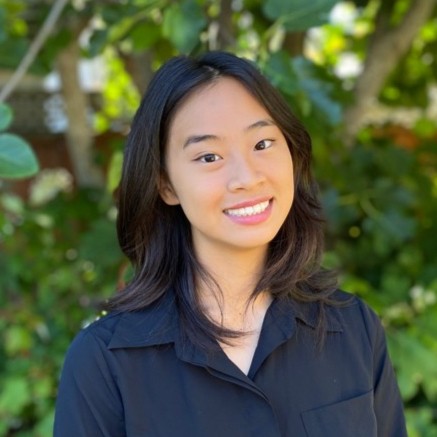}}]
{Kylie Lin} is a Ph.D. student in Human-Centered Computing at the Georgia Institute of Technology. In her research, she explores how visualization design decisions influence human cognition to identify strategies for effective visual data communication. Her recent work has leveraged modeling techniques to predict human perceptions of visual complexity based on visualization design features. %She has presented her work at venues across psychological and computer sciences, including IEEE VIS, ACM CHI, and Psychonomics.
\end{IEEEbiography}

\vspace{-1.2cm}

\begin{IEEEbiography}
[{\includegraphics[width=1in,height=1.25in,clip,keepaspectratio]
{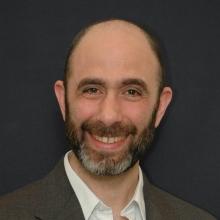}}]
{Andrew Cohen} is a Professor in the Cognition and Cognitive Neuroscience program in the Department of Psychological and Brain Sciences at the University of Massachusetts, Amherst. His research focuses on human reasoning and decision making, using both behavioral and computational modeling approaches.
\end{IEEEbiography}

\vspace{-1.2cm}

\begin{IEEEbiography}
[{\includegraphics[width=1in,height=1.25in,clip,keepaspectratio]
{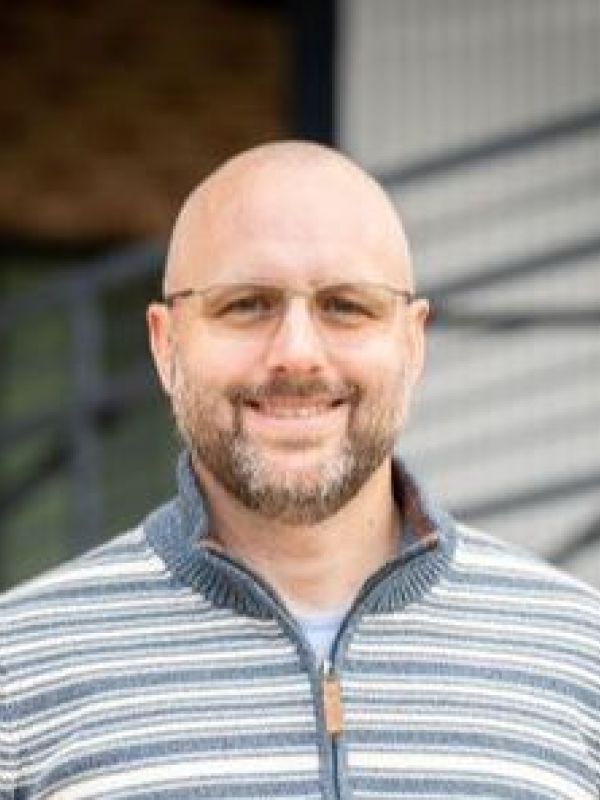}}]
{Ryan Kennedy} is a Professor in the College of Arts and Sciences Department of Political Science at Ohio State University. His research spans several areas in comparative politics, international relations, and American politics, including computational social science, technology and public policy, democratization, deliberation, energy and the environment, and prediction models.
\end{IEEEbiography}

\vspace{-1.2cm}

\begin{IEEEbiography}
[{\includegraphics[width=1in,height=1.25in,clip,keepaspectratio]
{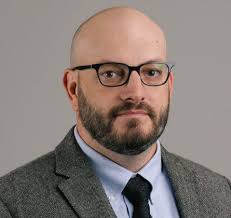}}]
{Zachary Zwald} is an Assistant Professor at the Department of Political Science at the University of Houston. His research examines judgment- and decision-making processes on issues at the intersection of technology and international security.  
\end{IEEEbiography}

\vspace{-1.2cm}

\begin{IEEEbiography}
[{\includegraphics[width=1in,height=1.25in,clip,keepaspectratio]
{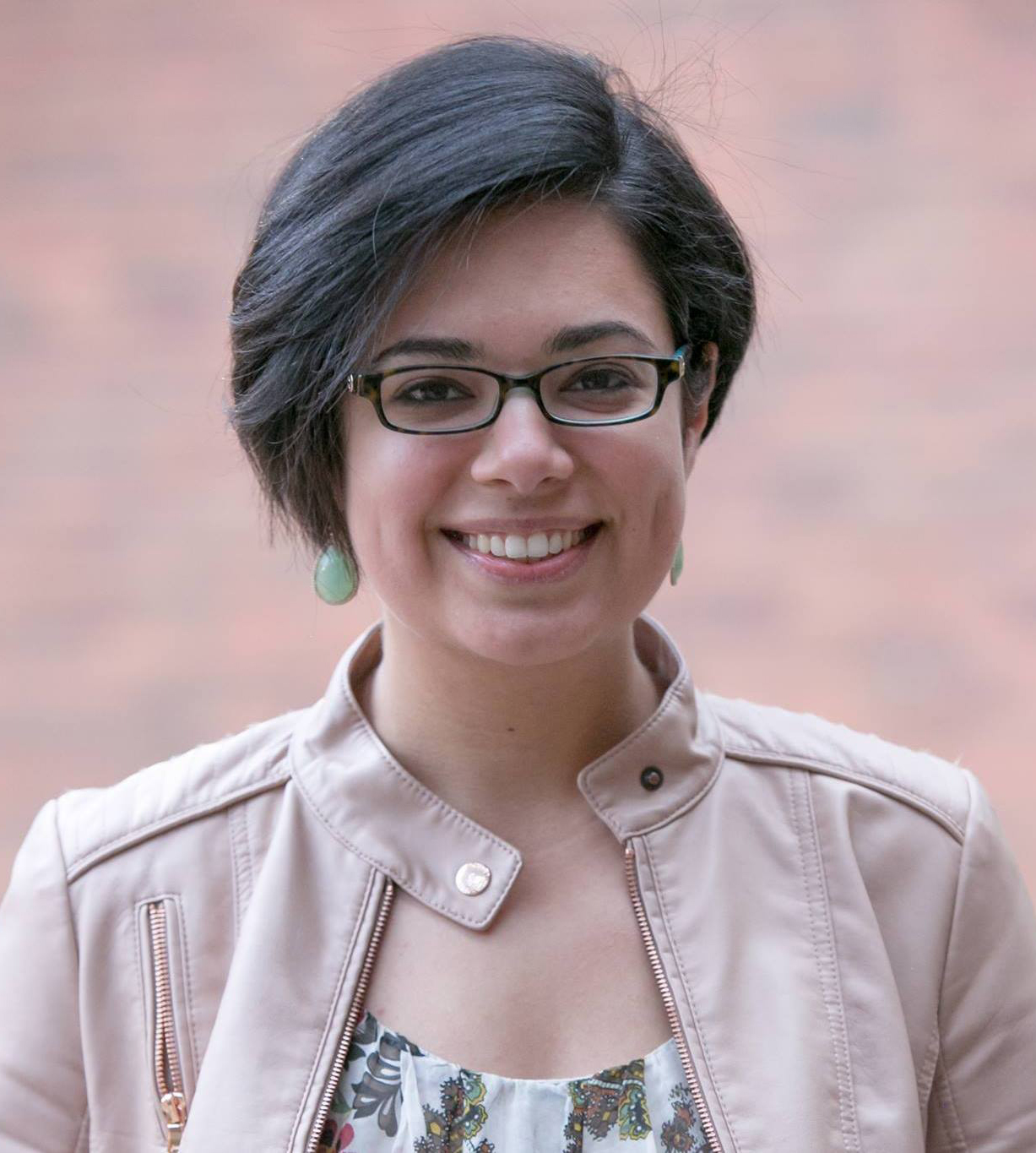}}]
{Carolina Nobre} is an Assistant Professor in the computer science department at the University of Toronto. Her research area is Data Visualization, focusing on creating user adaptive interactive data visualizations.
\end{IEEEbiography}

\vspace{-1.2cm}

\begin{IEEEbiography}
[{\includegraphics[width=1in,height=1.25in,clip,keepaspectratio]
{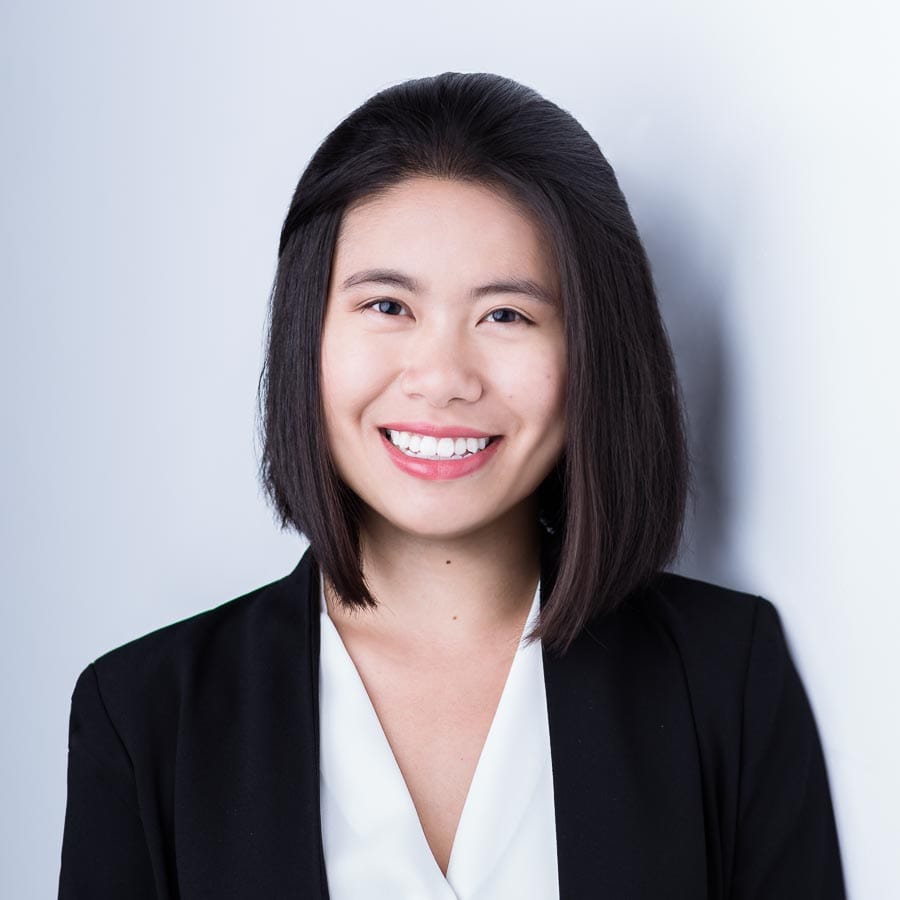}}]
{Cindy Xiong Bearfield} is an Assistant Professor in the School of Interactive Computing at the Georgia Institute of Technology. She received her Ph.D. in Cognitive Psychology and her MS in Statistics from Northwestern University.  Her research at the intersection of human perception, cognition, and data visualization has been recognized with an NSF CAREER award. She has received paper awards at premier psychology and data visualization venues, including ACM CHI, IEEE PacificVis, Psychonomics, and IEEE VIS.  
\end{IEEEbiography}

 

% \end{IEEEbiography}



\vfill

\end{document}